\documentclass[11pt,letterpaper,notoc]{JHEP3} 
\usepackage{epsfig}

\title{A non-abelian vortex lattice in strongly coupled systems}
\author{Kenny Wong  \\
Department of Applied Mathematics and Theoretical Physics, University of Cambridge, UK \\ 
{\tt k.wong@damtp.cam.ac.uk}
}

\abstract{The AdS/CFT correspondence predicts that background non-abelian magnetic fields induce instabilities in strongly-coupled systems with non-abelian global symmetries. These instabilities lead to the formation of vortex lattices in which the non-abelian currents ``antiscreen'' the applied magnetic field. From the bulk perspective, this behaviour can be traced to a well-known instability of Yang-Mills theory. We analyse the phase structure of the instability and comment on aspects of the vortex lattice.}

\begin{document}
\pagestyle{plain} \setcounter{page}{1}
\newcounter{bean}
\baselineskip16pt

\section{Introduction}

\paragraph{}
The purpose of this short note is to highlight a simple, yet striking, universal property of strongly coupled systems that is predicted by holography: an instability induced by a background magnetic field for a \emph{non-abelian} global symmetry.

\paragraph{}
Under this instability, the non-abelian current operators acquire vacuum expectation values, leading to the formation of a vortex lattice. In many respects, these vortices resemble the vortices in superfluids and superconductors. But there is a key difference: the vortex currents flow in the opposite direction! One may say that these vortices ``antiscreen'' the applied magnetic field. 

\paragraph{}
It is important to stress that the instability giving rise to these vortices is a property of systems at \emph{strong coupling}. As far as we are aware, there is no mechanism that drives an analogous instability in weakly-coupled theories with global non-abelian symmetries. Indeed, it is difficult to understand the origin of this effect without holography.

\paragraph{}
The fact that holography predicts an instability of this kind stems from an observation about Yang-Mills theory dating back to the 1980s:  non-abelian magnetic fields induce the formation of W-boson vortex lattices  \cite{ambjorn0, ambjorn1,ambjorn2}. These W-boson vortex lattices, implanted in AdS magnetic (or dyonic) black hole backgrounds, are dual to non-abelian vortex lattices in the strongly coupled boundary theory. W-boson vortices in AdS black holes were studied in some detail in \cite{erdmenger1, erdmenger2}, partly motivated by a proposal that abelian magnetic fields induce $\rho$-meson condensation \cite{chernodub1, chernodub2}. Here, however, we suggest a different interpretation: the instability is induced by a \emph{non-abelian} magnetic field in the boundary theory.

\paragraph{}
Of course, this begs the question: how can we realise non-abelian magnetic fields in condensed matter systems? Although the fundamental non-abelian fields of the Standard Model bear little relevence to condensed matter, it has been suggested that \emph{artificial} non-abelian magnetic fields can yet be realised in ultracold atomic gases \cite{ruseckas, osterloh, juzeliunas, satija2, hauke}. While holography is tenuously applicable to these particular setups, it is encouraging to know that non-abelian magnetic fields are experimentally feasible, providing hope that our generic prediction of AdS/CFT may be testable in the foreseeable future.

\paragraph{}
Throughout this note, we draw extensively upon the methods developed in previous studies of W-boson vortex lattices in AdS \cite{erdmenger1, erdmenger2}. Section 2 generalises the results of \cite{erdmenger1}, mapping out the phase diagram of the instability as one varies the magnetic field, chemical potential and temperature, for different values of the bulk coupling constants. Section 3 describes several aspects of the vortex lattice, including the antiscreening. We focus on the simpler $SU(2)$ case in Sections 2 and 3, reserving the generalisation to $SU(N)$ for the Appendix.

\section{The phase diagram of the instability}

\paragraph{}
Our system of interest is a planar, conformal, strongly coupled system with an $SU(2)$ global symmetry, which we model holographically as $SU(2)$ Yang-Mills theory in an asymptotically AdS$_4$ spacetime.
\begin{eqnarray}
S = \frac 1{2\kappa^2} \int d^4 x \sqrt{-g} \left(R+\frac 6 {L^2} \right) - \frac 1 {4e^2} \int d^4 x \sqrt{-g}    F_{\mu\nu}^a F^{\mu\nu a} \label{ymaction}
\end{eqnarray}
Physical properties of this theory only depend on the dimensionless ratio of coupling constants,
\begin{eqnarray}
\gamma = \frac{2 e^2 L^2 }{ \kappa^2}. \nonumber
\end{eqnarray}
$\gamma$ is a measure of the total number of degrees of freedom in the boundary theory, divided by the number of degrees of freedom charged under the $SU(2)$ global symmetry.

\paragraph{}
The boundary theory has three adjustable parameters. Naturally, one of them is the background non-abelian magnetic field $B$. We also allow for a non-abelian chemical potential $\mu$ in the same component of the Lie algebra. These appear in the bulk as sources for the $A_\mu^3$ field,
\begin{eqnarray}
A_y^3 \to Bx, \qquad A_t^3 \to -\mu \qquad \qquad z \to 0.
\end{eqnarray}
The final parameter is the temperature $T$ of the boundary theory, which sets the Hawking temperature of the bulk horizon.

\paragraph{}
In \cite{ambjorn1,ambjorn2}, it was shown that a magnetic field $F_{xy}^3 = Bx$ induces an instability of Yang-Mills theory in flat space: the magnetic field causes the off-diagonal ``W-bosons'' $W_\mu = A_\mu^1 - iA_\mu^2 $ to become tachyonic and condense. The magnetic field $F_{xy}^3$ breaks the $SU(2)$ gauge group to a diagonal $U(1)$ subgroup; this diagonal $U(1)$ subgroup is subsequently broken spontaneously when the W-bosons condense.

\paragraph{}
This magnetically-induced W-boson instability was studied in AdS$_5$ black holes in \cite{erdmenger1}. A new feature that arises in a black hole background is that the Hawking temperature $T$ determines the critical magnetic field $B_c$ at which the W-bosons become unstable.

\paragraph{}
The goal of this section is to explore the full parameter space of the model and determine the conditions under which the W-bosons condense. Along the way, we make two counterintuitive observations:
\begin{itemize}
\item W-bosons in AdS$_4$ are only susceptible to the instability if $\gamma > \frac 3 4$. For the strongly coupled system on the boundary, this means that non-abelian magnetic fields will only induce the instability if the charged degrees of freedom are sufficiently dilute, that is, if the total number of degrees of freedom sufficiently exceeds the number of charged degrees of freedom. This result follows naturally from our holographic analysis in the bulk, but is difficult to rationalise from the perspective of the boundary theory.
\item Provided that $ \frac 3 4 < \gamma < 3$, turning on a non-abelian chemical potential source in AdS$_4$ (while holding the non-abelian magnetic field source fixed) has the effect of \emph{suppressing} the formation of a W-boson condensate. This result  goes against conventional wisdom about holographic superconductors, where chemical potentials usually induce, rather than suppress, the formation of a condensate that breaks gauge symmetry.
\end{itemize}

\paragraph{}
The magnetically-induced instability is sensitive to the values of the parameters $B, \mu, T$ and $\gamma$ because they determine the form of the background. All four of these parameters appear in the bulk metric\footnote{Reference \cite{erdmenger1} takes the probe limit $\gamma \to \infty$ in AdS$_5$ and sets $\mu = 0$. We allow $\mu \geq 0$ and focus predominantly on understanding the properties of the theory at finite $\gamma$. We work in AdS$_4$, where the black hole metric is known explicitly for finite $\gamma$.},
\begin{eqnarray}
ds^2 = \frac 1 {z^2} \left( -f(z) dt^2 + \frac { dz^2} { f(z)} + dx^2 + dy^2 \right). \label{bggeom}
\end{eqnarray}
The emblackening factor is
\begin{eqnarray}
f(z) = 1 - (4 - 4\pi T z_h ) \frac{z^3}{z_h^3} + (3-4\pi T z_h) \frac{z^4}{z_h^4}, \nonumber
\end{eqnarray}
and the position of the horizon can be found by solving
\begin{eqnarray}
4\pi Tz_h = 3 - \frac 1 \gamma \left( B^2 z_h^4 + \mu^2 z_h^2 \right). \label{horposition}
\end{eqnarray}
The parameters also determine the background gauge field,
\begin{eqnarray}
A_y^3 = Bx, \qquad A_t^3 = -\mu (1 - z/z_h).  \label{bggauge}
\end{eqnarray}

\paragraph{}
Of course, (\ref{bggeom}) and (\ref{bggauge}), together with $W_\mu = 0$, is a solution to (\ref{ymaction}) that preserves the diagonal $U(1) \subset SU(2)$ gauge symmetry. However, we shall see that, under certain conditions on $B, \mu, T$ and $\gamma$, the W-bosons $W_\mu$ are unstable in this background, and will condense and spontaneously break this $U(1)$ symmetry. Our task is to determine what these conditions are.

\paragraph{}
To perform the instability analysis, we must study the Yang-Mills equations for different values of the parameters. The Yang-Mills equations are non-linear and to solve them is a difficult task. But for the purposes of locating the \emph{onset} of the instability, it is sufficient to expand the Yang-Mills equations about the background (\ref{bggeom}), (\ref{bggauge}) to linear order in the W-boson fields \cite{erdmenger1}. We set $W_z = 0$ by a choice of gauge, and we are also free to set $W_t = 0$ since the $W_t$  component is stable\footnote{The linearised equation of motion for $W_t$ is $(f(z) \partial_z^2 + \partial_x^2  +D_y^2) W_t - i\mu (1 - z/z_h) (\partial_x W_x + D_y W_y) = 0$. Substituting $W_t = \tilde W_t (z) e^{-iky} e^{-\frac B 2 (x + \frac k B)^2 }$ and equation (\ref{Wmodes}) for the lowest Landau level, this reduces to $f(z) \partial_z^2 \tilde W_t - B \tilde W_t = 0$, from which it is clear that $\tilde W_t$ acquires a positive square mass from its interaction with the magnetic field. Thus $W_t$ is stable. One can also neglect the equations of motion for $A_\mu^3$ and $g_{\mu\nu}$ in our linearised analysis because they are unsourced by $W_x$ and $W_y$ at linear order.}. At linear order in $W_\mu$, the equations of motion are
\begin{eqnarray}
\partial_z ( f(z) \partial_z W_x ) + D_y (D_y W_x - \partial_x W_y ) + f(z)^{-1} \mu^2 (1-z/z_h)^2 W_x  - iB W_y &= &0 \nonumber \\
\partial_z ( f(z) \partial_z W_y ) + \partial_x (\partial_x W_y -D_y W_x ) + f(z)^{-1} \mu^2 (1-z/z_h)^2 W_y+ iB W_x &=& 0 \nonumber \\
\partial_z (\partial_x W_x + D_y W_y ) &=& 0 \label{Wxyeqns}
\end{eqnarray}
with $D_y = \partial_y - iBx$.

\paragraph{}
Only the lowest Landau level  modes of the W boson field are susceptible to the instability. These modes take the form
\begin{eqnarray}
W_x = -i W_y = \tilde W(z) e^{-iky} e^{-\frac B 2 (x + \frac k B)^2 } \label{Wmodes}
\end{eqnarray}
where $k$ is a real parameter, and where the radial profile function $\tilde W(z)$ obeys
\begin{eqnarray}
\partial_z (f(z) \partial_z \tilde W(z) ) + f(z)^{-1} \mu^2 (1-z/z_h)^2 \tilde W(z) + B \tilde W(z) = 0. \label{radial}
\end{eqnarray}
It is evident from (\ref{radial}) that the lowest Landau level modes are tachyonic: they acquire a negative square mass through their interaction with the magnetic field.

\paragraph{}
Note that, if we set $B = 0, \mu > 0$, our model reduces to the standard holographic model of a $p$-wave superconductor \cite{gubser}. Clearly, the chemical potential term in (\ref{radial}) gives $\tilde W(z)$ a negative square mass; this is precisely the origin of the superconducting instability in the holographic $p$-wave superconductor model. We do not, however, interpret our model as a holographic superconductor here: while magnetic fields have occasionally been observed to induce superconductivity \cite{uji, levy}, it is much more the norm for superconductivity to be suppressed by magnetic fields in the laboratory.

\subsubsection*{\underline{$\mu = 0$, $T = 0$:} Instabilities for dilute charges only}

\paragraph{}
In flat space, the existence of a tachyonic mode guarantees an instability \cite{ambjorn1}. But tachyonic modes in AdS are not always unstable; they only become unstable when they violate their Breitenlohner-Freedman bound. At zero temperature, the instability of our theory is driven by the violation of the BF bound in the asymptotically AdS$_2 \times \mathbb R^2$ region near the IR horizon, similar to the familiar instability in the $s$-wave holographic superconductor \cite{holosuper}.

\paragraph{}
To determine the conditions for an instability to occur, let us therefore write the radial equation of motion (\ref{radial}) in terms of the AdS$_2\times \mathbb R^2$ radial coordinate $\zeta = z_h^2 / 6(z_h - z)$, starting with the case where $\mu = 0$. Using (\ref{horposition}) to express the position of the extremal horizon in terms of $\gamma$, equation (\ref{radial}) reduces in the near-horizon limit to
\begin{eqnarray}
\partial_\zeta^2 \tilde W + \frac{\gamma}{\sqrt{12}} \frac {\tilde W}{\zeta^2 } = 0.
\end{eqnarray}
This has solutions of the form $\tilde W \sim \zeta^{\Delta_\pm}$, and the exponents $\Delta_\pm$ are complex when
\begin{eqnarray}
\gamma > \frac 3 4. \label{Bonly}
\end{eqnarray}
When this inequality holds, the W-bosons violate their BF bound in the IR region of the spacetime. Therefore, when $\gamma > \frac 3 4$, the W-bosons condense at zero temperature in the presence of a non-abelian magnetic field. 

\paragraph{}
From the perspective of the bulk, $\gamma$ describes the relative strengths of the gauge and gravitational forces. On the one hand, the magnetic field lowers the square mass of the W-boson via the gauge interaction, making the lowest Landau level of the W-boson more tachyonic. But on the other hand, the magnetic field also relaxes the BF bound through its effect on the gravitational metric. The value of $\gamma$ determines which of the two effects dominates.

\paragraph{}
From the point of view of the dual boundary theory, the $\gamma > \frac 3 4$ condition has a surprising interpretation: it is a bound on the ratio between the total and charged degrees of freedom. ($\gamma$ has a similar interpretation in conventional phenomenological models of holographic superconductors \cite{holosuper}.) The number of charged degrees of freedom must be sufficiently \emph{low} for the instability to occur; in other words, the system must contain a sufficient amount of neutral matter. This result is clear from holography but it is difficult it to understand directly from the perspective of the boundary theory.

\subsubsection*{ \underline{$\mu > 0$, $T = 0$:} Instability inhibited by chemical potential}

\begin{figure}[!h]
\begin{center}
\includegraphics[trim = 0.0in 0.6in 0.0in 0.0in, width=2.8in, height = 1.8in]{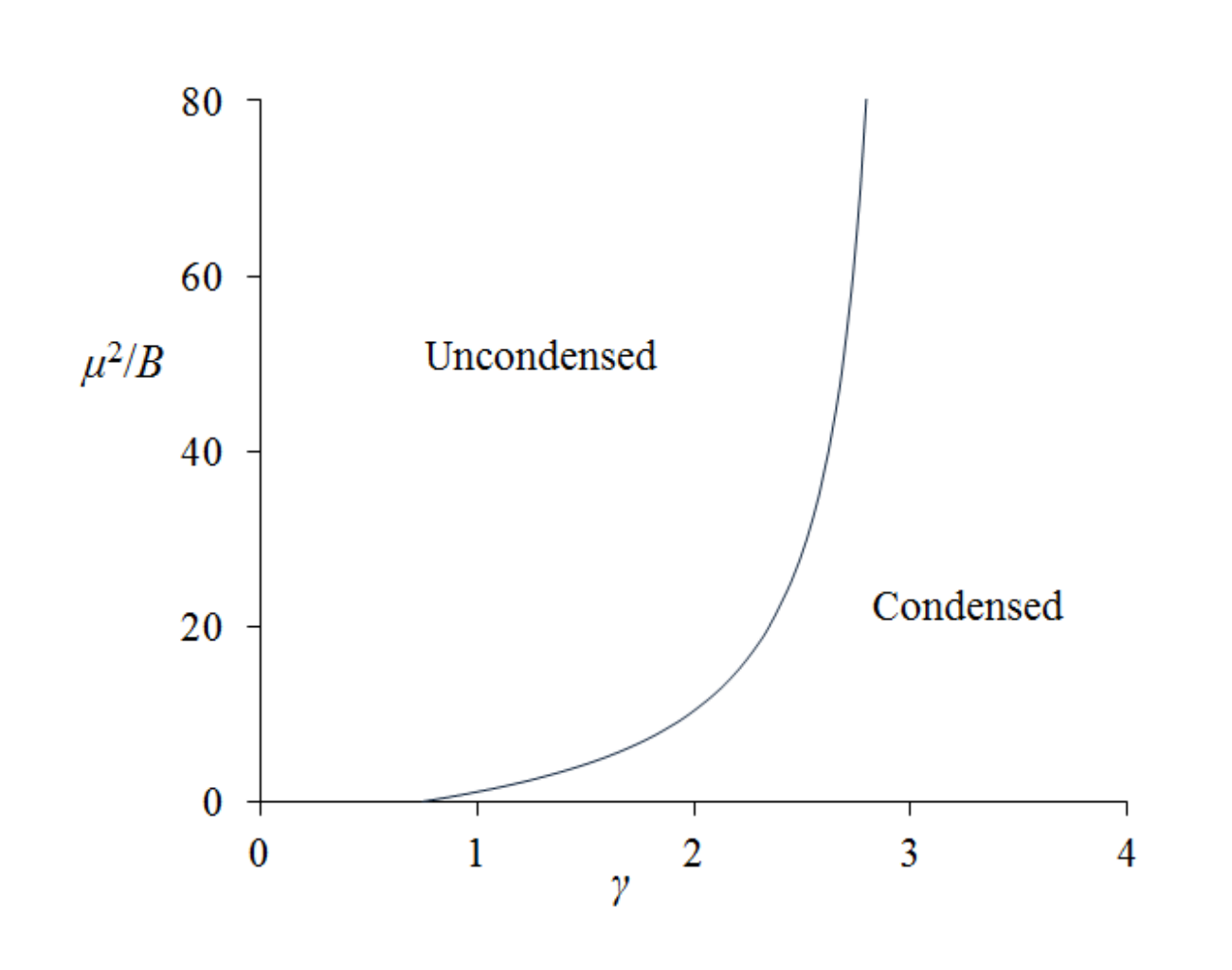}
\end{center}
\caption{Phase diagram at zero temperature}
\label{Dyonzero}
\end{figure}

\paragraph{}
The above analysis can be repeated with $B$ and $\mu$ present simultaneously. The condition for instability at zero temperature becomes
\begin{eqnarray}
\gamma > \frac 3 4 \left(\frac{1+\mu^2 /3B}{1+\mu^2 /6B} \right)^2 . \label{Bandmu}
\end{eqnarray}
This result is illustrated in Figure \ref{Dyonzero}.

\paragraph{}
Two limiting cases are of interest. In the limit $\mu \to 0$, we recover our previous result, $\gamma > \frac 3 4$. In the limit $B \to 0$, in which our model reduces to the standard holographic model of a $p$-wave superconductor, we recover the well-known condition $\gamma > 3$ that holds ubiquitously in the theory of holographic superconductors \cite{holosuper}.

\paragraph{}
The fact that the critical values of $\gamma$ differ by a factor of four in these two limiting cases leads to an interesting observation. When $\frac 3 4 < \gamma < 3$, the W-bosons are condensed when $B > 0$ and $\mu = 0$. But as $\mu$ is increased with $B$ held constant, the system returns to the uncondensed phase; that is to say, \emph{raising the chemical potential has the effect of destroying the condensate}. This kind of behaviour is rather unexpected in the context of holographic superconductors, where chemical potential sources usually act in favour of the instability.

\subsubsection*{\underline{$\mu > 0$, $T > 0$:} Full phase diagram}

\begin{figure}[!h]
\begin{center}
\includegraphics[trim = 0.0in 0.4in 0.0in 0.2in, width=6.4in, height = 1.9in]{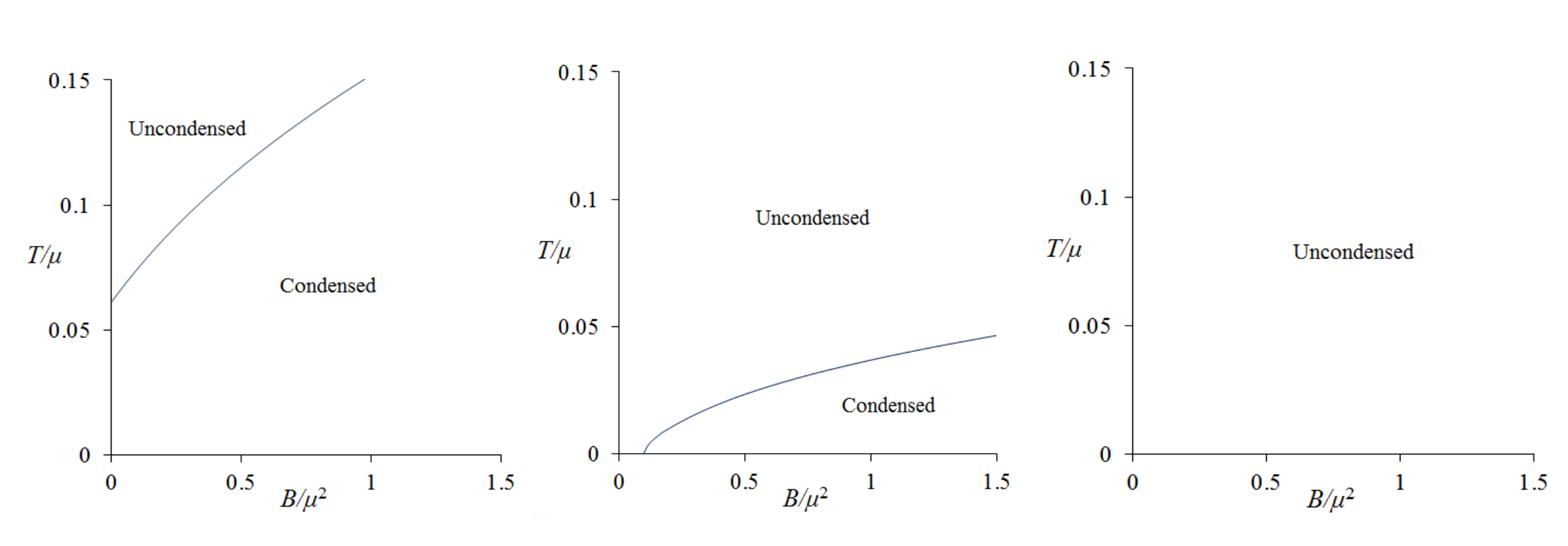}
\end{center}
\caption{Phase diagram for $\gamma \to \infty$ (left), $\gamma = 2$ (middle),  $\gamma = 0$ (right)}
\label{Dyonprobe}
\end{figure}

\paragraph{}
Any condensates that do form at zero temperature are destroyed as the temperature is increased. The critical temperature at which this occurs can be determined numerically: for any given $\gamma$ and $\mu^2 / B$, one simply computes the lowest value of $T/\sqrt B$ at which there exists a normalisable solution to (\ref{radial})\footnote{We solve (\ref{radial}) by 4th order Runge-Kutta. We impose Dirichlet boundary conditions in the UV and regularity in the IR using the shooting method. The numerics are unreliable for $T \ll \sqrt B$, and for this reason, Figure \ref{Dyonprobe} interpolates numerical results for $ T \gg \sqrt B$ with the analytic result (\ref{Bandmu}) for $T = 0$.}. Typical results of this computation are plotted in Figure \ref{Dyonprobe}.

\paragraph{}
One can identify three qualitatively distinct ranges of $\gamma$:
\begin{itemize}
\item $\gamma >3$ (left): The system is in the condensed phase at zero temperature, but the condensate is destroyed if the temperature is raised above a certain critical temperature $T_c$. The value of $T_c$ increases as $B$ increases.
\item $ \frac 3 4 < \gamma <  3$ (middle): If $B / \mu^2$ is sufficiently large, the system is in the condensed phase at zero temperature. Again, $T_c$ increases as $B$ increases.
\item $ \gamma < \frac 3 4$ (right): The system is in the uncondensed phase throughout.
\end{itemize}

\paragraph{}
In the limit $\gamma \to \infty$ (Figure \ref{Dyonprobe}, left), the geometry reduces to AdS$_4$-Schwarzschild. Here, it is instructive to compare our numerical results with analytical estimates obtained using a WKB method\footnote{We adapt the method of \cite{erdmenger1} to four spacetime dimensions. The radial equation (\ref{radial}) can be written in Schrodinger form, $-\partial_{\tilde z}^2 \tilde W + V(\tilde z) \tilde W = 0$, where  $\tilde z = \int dz / f(z) \in (0,\infty) $ is the tortoise coordinate and $V(\tilde z) = - B f(\tilde z)$ is the potential. $V(\tilde z)$ tends to  zero continuously in the IR but is cut off discontinuously in the UV by the Dirichlet boundary condition. It follows by a standard WKB argument that there exist zero-energy bound states when $\int_0^\infty \sqrt{-V(\tilde z)} d \tilde z \approx (n+ \frac 3 4 ) \pi$. Seeking an $n=0$ bound state, we obtain $T_c\approx 3^{-3/2}\pi^{-1} \mu$ for $B= 0$ and $T_c \approx  \pi^{- 3/ 2}  \Gamma(\frac 4 3)  \Gamma(\frac 5 6)^{-1}  B^{1/2} $ for $B \gg \mu^2$.} adapted from \cite{erdmenger1}. The numerical and analytic results agree to within 1 percent.

\subsubsection*{$p$ versus $p+ip$}

\paragraph{}
When $B > 0$, the unstable mode (\ref{Wmodes}) has the property $W_x = -iW_y$. This observation was made in the context of Yang-Mills theory in flat space \cite{ambjorn0}. From the point of view of holography, the relationship $W_x = -iW_y$ indicates that the condensate that forms in the boundary theory is a $(p+ip)$-wave condensate. (By contrast, a $p$-wave condensate ($W_y = 0$) cannot form at the onset of instability because a lowest Landau level ansatz of the form $W_x  = \tilde W(z) e^{-iky} e^{-\frac B 2 (x + \frac k B)^2 }$ and $W_y = 0$ does not obey the linearised equations of motion.)

\paragraph{}
It is interesting to observe that this is markedly different from the condensates that form when $B = 0,$ $\mu > 0$. There, the $p$-wave and $(p+ip)$-wave instabilities are both solutions of the linearised equations of motion, so it is not possible to discern which of the two is favoured by doing a linearised analysis alone. To determine the preferred ground state in this case, one has to compare the free energies of the two types of condensates. This analysis was performed in \cite{gubser}, where it was shown that the $p$-wave condensate is energetically favourable when $B = 0,$ $\mu > 0$.

\section{Non-abelian antiscreening vortices}

\paragraph{}
The W-boson condensate induced by this instability of Yang-Mills takes the form of a vortex lattice \cite{ambjorn2}. The purpose of this section is to discuss the holographic interpretation of the AdS$_4$ W-boson vortex lattice, using the techniques developed in \cite{erdmenger2}.

\paragraph{}
Unsurprisingly, the holographic dual of a vortex lattice is a vortex lattice. What is more surprising is that the currents $\vec{j}^a$ point in the opposite direction relative to the vortex currents in conventional superfluids and superconductors.

\paragraph{}
One can characterise the direction of the flow by comparing the vorticity field $\vec{\omega}^a = \nabla \times \vec{j}^a$ to the applied magnetic field $\vec B^a$ at different positions. In our strongly coupled system, the product $\vec{\omega}^a . \vec B^a$ is negative at the vortex cores, and is positive in the spaces between the vortices. This is precisely the reverse of the flow pattern observed in superconductors. It is also the reverse of the flow pattern observed in superfluids, where the applied angular velocity substitutes for the applied magnetic field.

\paragraph{}
Alternatively, one can say that the vortices in our system ``antiscreen'' the applied magnetic field: if the $SU(2)$ global symmetry of the boundary theory is weakly gauged and the magnetic field is made dynamical, then the magnetic field induced by the vortex currents would \emph{enhance} the applied magnetic field  in the regions between the vortices. In the context of Yang-Mills theory in flat space, this phenomenon was described as the ``anti-Lenz'' effect \cite{ambjorn1,ambjorn2}, and was shown to be a consequence of asymptotic freedom \cite{ambjorn0}.

\paragraph{}
We begin our derivation of this result by reviewing the construction of the W-boson lattice. The solution is not known analytically, but it was observed in \cite{erdmenger2} that solutions to the linearised Yang-Mills equations are good approximations to the full solution when $T$ is close to $T_c$. Taking a suitable linear combination of the solutions (\ref{Wmodes}) to our linearised equations (\ref{Wxyeqns}), we can form a class of rhombic lattices,
\begin{eqnarray}
W_x = -iW_y = \tilde W(z) \sum_{n = -\infty}^{\infty} \exp \left( - \frac{i\pi n^2} 2 - \frac{2\pi i \sqrt{B} n y} \lambda - \frac B 2 \left( x + \frac{2\pi n}{\lambda \sqrt{B}} \right)^2 \right). \label{lattice}
\end{eqnarray}
The parameter $\lambda \in \mathbf R$ determines the shape of the lattice. Up to a gauge transformation, these lattices are invariant under the translations
\begin{eqnarray}
(x,y)\mapsto (x, y+ \lambda /\sqrt B) \qquad (x,y) \mapsto (x + 2\pi/ \lambda \sqrt B , y + \lambda / 2\sqrt B).
\end{eqnarray}
One must go beyond the linearised approximation to determine the value of $\lambda$ that minimises the free energy \cite{erdmenger2}\footnote{In  \cite{erdmenger2}, it is shown that the triangular lattice ($\lambda = \sqrt{4\pi} / 3^{1/4}$) minimises the free energy in a five-dimensional version of this model in the limit $\gamma \to \infty$. However, the difference in energy between the triangular lattice ($\lambda = \sqrt{4\pi} / 3^{1/4}$) and the square lattice ($\lambda = \sqrt{4\pi} $) in \cite{erdmenger2} is only of the order of one percent, and is therefore likely to be offset by any small deformations that one may make to the bulk model. Since $\lambda$ is susceptible to change under small deformations of the bulk model, we leave $\lambda$ as an undetermined parameter throughout this paper.}. Similarly, the overall normalisation of $\tilde W(z)$ can only be determined by examining the equations of motion beyond linear order \cite{erdmenger2}, although, as we will see later, it is possible to extract how this normalisation depends on temperature without solving the equations explicitly.

\subsubsection*{\underline{$\mu = 0$:} Antiscreening vortices}

\begin{figure}[!h]
\begin{center}
\includegraphics[trim = 0.5in 0.53in 0.5in 0.13in, width=5.in, height = 2.5in]{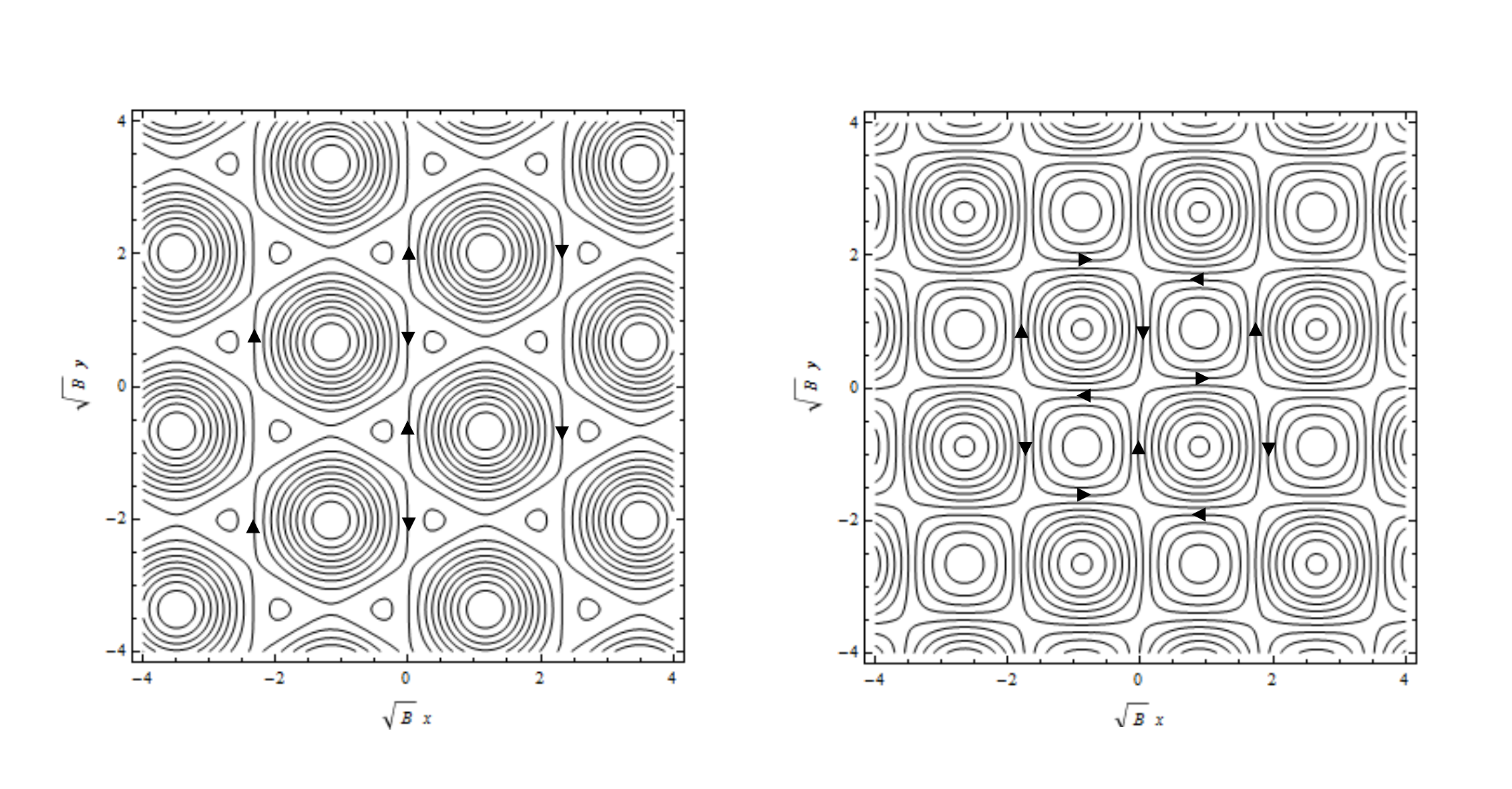}
\end{center}
\caption{Non-abelian vortices in the boundary theory for $\lambda = \sqrt{4\pi}/3^{1/4}$ (left) and $\lambda = \sqrt{4\pi}$ (right). The applied non-abelian magnetic field points out of the page.}
\label{Currents}
\end{figure}

\paragraph{}
To compute the vortices in the boundary theory, one must calculate $A_\mu^3$ up to quadratic order in $W_\mu$. The subleading fall-offs of $A_\mu^3$ encode the expectation values of the non-abelian current operator $j_\mu^3$,
\begin{eqnarray}
A_x^3 \sim e^2 z \langle j_x^3 \rangle, \qquad A_y^3 \sim Bx + e^2 z \langle j_y^3 \rangle, \qquad \qquad z \to 0.
\end{eqnarray}
Throughout this section, we work in the $\gamma \to \infty$ probe limit, where we may ignore corrections to the metric. Using $a_\mu^3$ to denote corrections  to the background gauge field (\ref{bggauge}) of quadratic order in $W_\mu$, the Yang-Mills equations are solved to quadratic order in $W_\mu$ by
\begin{eqnarray}
a_x^3 = +\partial_y \chi, \qquad a_y^3 = -\partial_x \chi, \qquad a_z^3=a_t^3 = 0,
\end{eqnarray}
where $\chi(x,y,z)$ is a function that satisfies
\begin{eqnarray}
\partial_z (f(z) \partial_z \chi )+ \partial_x^2 \chi + \partial_y^2 \chi =  -  \vert W_x \vert^2.  \label{chieqn}
\end{eqnarray}
$\langle j_x^3 \rangle$ and $\langle j_y^3 \rangle$ can be obtained by solving (\ref{chieqn}) numerically, following the method of \cite{erdmenger2}\footnote{$\tilde W(z)$ is first obtained numerically from (\ref{radial}). Decomposing $\vert W_x \vert^2 $ and $a_\mu$ into Fourier modes, one then reduces (\ref{chieqn}) to an infinite set of decoupled ODEs, of which only the lowest Fourier modes contribute significantly to the final result. These ODEs are solved numerically subject to Dirichlet boundary conditions in the UV and regularity in the IR.}. The result of this computation is plotted in Figure \ref{Currents}, which shows the antiscreening behaviour: the currents $\vec{j}^3$ flow clockwise around the vortex cores while the applied magnetic field $\vec{B}^3$ points out of the page.

\subsubsection*{\underline{$\mu > 0$:} Charge density modulations}

\begin{figure}[!h]
\begin{center}
\includegraphics[trim = 0.5in 0.53in 0.5in 0.13in, width=5.in, height = 2.5in]{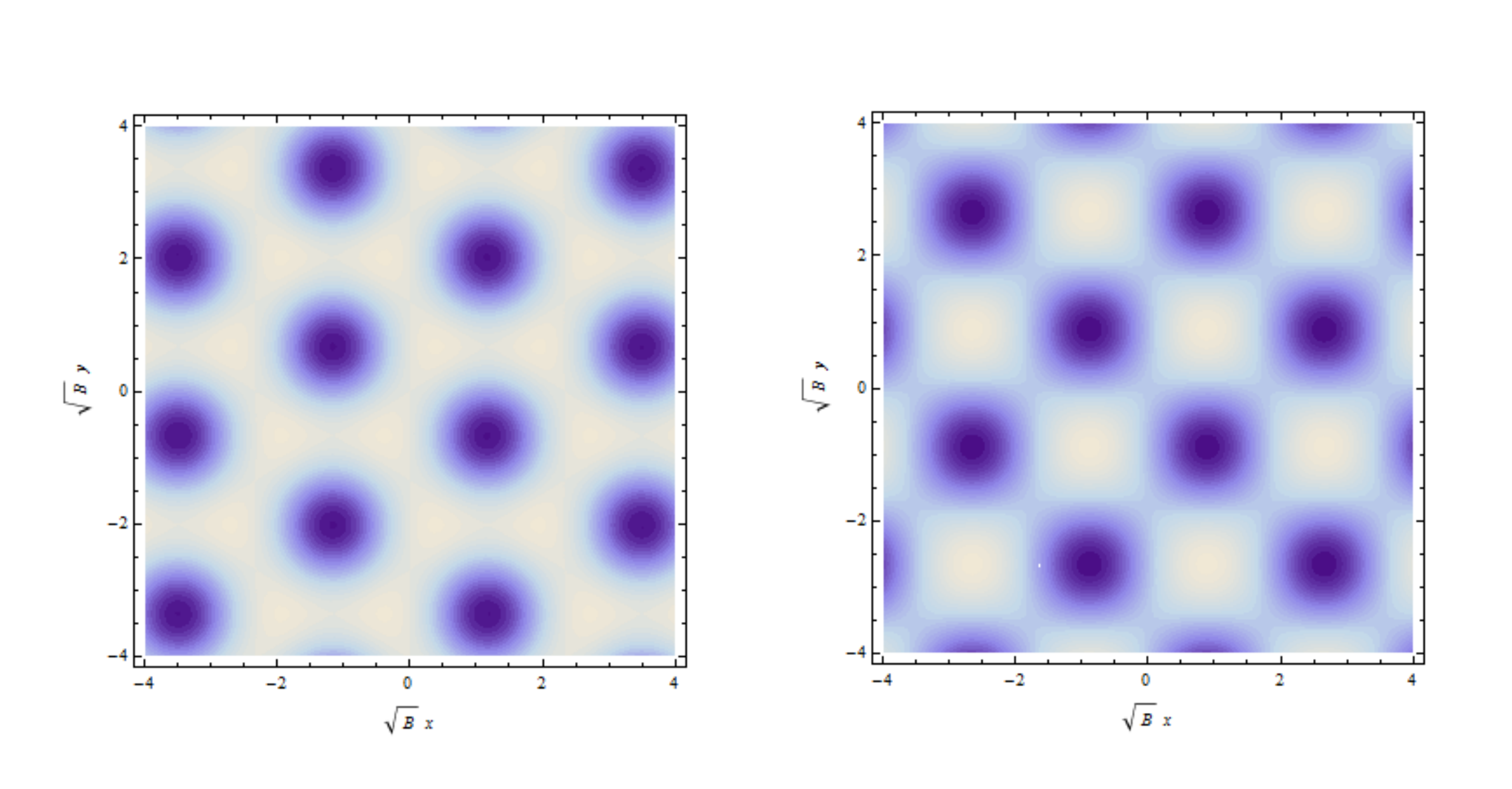}
\end{center}
\caption{Non-abelian charge density modulations in the boundary theory for $\lambda = \sqrt{4\pi}/3^{1/4}$ (left) and $\lambda = \sqrt{4\pi}$ (right). Darker shading indicates lower values of charge density.}
\label{Charge}
\end{figure}

\paragraph{}
One can perform a similar calculation in the presence of a non-abelian chemical potential source. In addition to a vortex lattice similar to what is shown in Figure \ref{Currents}, one finds that the boundary theory also acquires a spatially-modulated distribution of non-abelian charge density $j_t^3$.

\paragraph{}
The charge density of the boundary theory is given by the subleading fall-off of $A_t^3$,
\begin{eqnarray}
A_t^3 \sim -\mu +  e^2 z \langle j_t^3 \rangle, \qquad \qquad z \to 0. \nonumber
\end{eqnarray}
Writing the quadratic-order correction to $A_t^3$ as $a_t^3$, one determines $\langle j_t^3 \rangle$ by solving
\begin{eqnarray}
f(z) \partial_z^2 a_t^3 + \partial_x^2 a_t^3 + \partial_y^2 a_t^3 & = & - 2\mu (1- z/z_h) \vert W_x \vert^2. \label{ateqn}
\end{eqnarray}
The result of this calculation is illustrated in the limit $\mu \ll \sqrt B$ in Figure \ref{Charge}.  The background gauge field (\ref{bggauge}) contributes a constant non-abelian charge density $\langle j_t^3 \rangle = \mu/e^2z_h$, and on top of this, the quadratic-order correction $a_t^3$ adds a spatially-modulated contribution. The minima of the charge density distribution are situated at the vortex cores.

\subsubsection*{Scaling of current and charge density}

\paragraph{}
With $B$ fixed, the magnitudes of the vortex currents and charge density modulations in the boundary theory are proportional to $T_c - T$ near the onset of the instability. This follows from the fact that the normalisation of $\tilde W(z)$ is proportional\footnote{The scaling of $\tilde W(z)$ does not merely follow from dimensional analysis, but it can be determined by examining the equation for $W_\mu$. Using $W_{\mu,{\rm Lin}}$ to denote the solution to the linearised equation (\ref{Wxyeqns}) and using $w_\mu$ to denote its cubic-order correction, the leading-order corrections to (\ref{Wxyeqns}) take the schematic form
\begin{eqnarray}
\vec{\mathcal D} w_\mu  = \vec{\mathcal L} \left(T_c - T, a_\mu, \vert W_{{\rm Lin}} \vert^2 \right) W_{\mu, {\rm   Lin}}. \label{wthird}
\end{eqnarray}
Here, one has used the relation $z_h = 3/4\pi T$ to rewrite the equation for $W_\mu$ in terms of a dimensionless coordinate $\bar z = z / z_h$, so that $T$ now appears explicitly as a parameter in the equations. $\vec{\mathcal D}$ and $\vec{\mathcal L}$ are position-dependent linear operators, and $\mathcal L$ depends linearly on $T_c - T$, $a_\mu$ and $\vert W_{{\rm Lin}}\vert^2$. Equation (\ref{wthird}) must be solved subject to Dirichlet boundary conditions in the UV, regularity in the IR and the further requirement that $\vert w_\mu \vert \sim \mathcal O \left(\vert W_{{\rm Lin}}\vert^3 \right) $.

Now suppose that, for a given $\vert W_{{\rm Lin}}\vert$ and for a given $T$, there exists a solution $w_\mu$ to (\ref{wthird}) satisfying all of the above conditions. By linearity, there will continue to exist a solution to (\ref{wthird}) as $T$ is increased, as long as the normalisation of $\vert W_{{\rm Lin}} \vert$ increases in proportion to $(T_c - T)^{1/2}$. (Note that (\ref{chieqn}) and (\ref{ateqn}) guarantee that $a_\mu$ increases in proportion to $T_c - T$ when $\vert W_{{\rm Lin}} \vert$ increases in proportion to $(T_c - T)^{1/2}$.)} to $ (T_c - T)^{1/2}$.

\acknowledgments
The author is grateful to David Tong for supervision of this work and for helpful comments on drafts of this preprint. The author also acknowledges useful discussions with Benjamin Beri, Mike Blake, David Khmelnitskii and Michal Kwasigroch. The author is supported by the ERC STG grant 279943, ``Strongly Coupled Systems".

\appendix
\section{Appendix: $SU(N)$}

\paragraph{}
Our instability analysis generalises in a simple way to $SU(N)$ symmetry groups. We use a Cartan-Weyl basis for $\mathfrak{su}(N)^{\mathbb{C}}$,
\begin{eqnarray}
[H_i, H_j] = 0 \qquad [H_i, E_{\vec\alpha} ] = \alpha_i E_{\vec\alpha} && \qquad [E_{\vec\alpha}, E_{-\vec\alpha} ] = \alpha_i H_i \qquad [E_{\vec\alpha}, E_{\vec\beta} ] = N_{\vec\alpha \vec\beta} E_{\vec\alpha + \vec\beta}, \nonumber \\
{\rm tr}(H_iH_j) = \frac 1 2 \delta_{ij} \qquad {\rm tr}(H_i  E_{\vec\alpha} ) = 0&& \qquad {\rm tr}(E_{\vec\alpha}  E_{-\vec\alpha}) = \frac 1 2 \qquad {\rm tr}(E_{\vec\alpha} E_{\vec\beta}) = 0. \nonumber
\end{eqnarray}
We denote the gauge field components by $A_\mu = A_{i\mu} H_i + \frac 1 {\sqrt 2} \sum_{\vec\alpha} W_{\vec\alpha \mu} E_{\vec\alpha} $ and impose the reality constraint $A_{i\mu}^\dagger = A_{i\mu},  W_{\vec\alpha\mu}^\dagger = W_{-\vec\alpha\mu}$.

\paragraph{}
For $SU(N)$, sources may be introduced for any component of the Cartan subalgebra,
\begin{eqnarray}
A_y \to B_i x H_i , \qquad A_t \to -\mu_i H_i , \qquad \qquad z \to 0. \nonumber
\end{eqnarray}
The sources explicitly break $SU(N)$ to the $U(1)^{N-1}$ subgroup generated by elements of the Cartan subalgebra. (Further symmetries are preserved if $B_i$ and $\mu_i$ lie in Weyl planes.)

\paragraph{}
Should a W-boson component $W_{\vec \alpha \mu}$ condense, the $U(1)$ subgroup generated by $\alpha_i H_i$ is spontaneously broken. This gives rise to expectation values for current and charge density operators corresponding to this generator in the dual boundary theory.

\paragraph{}
Repeating the calculation in the main text, one finds that the $W_{\vec\alpha\mu}$ gauge field component is unstable at zero temperature if
\begin{eqnarray}
\gamma >  \frac{ 27B_i B_i + 18\mu_i \mu_i B_j \alpha_j + 3 \mu_i \mu_i (\mu_j \alpha_j)^2}{\left( 6B_i \alpha_i + (\mu_i \alpha_i )^2 \right)^2}. \nonumber
\end{eqnarray}
When $\mu_i = 0$, this inequality reduces to $\gamma > \frac 3 4( B_i B_i )/ (B_j\alpha_j)^2 $, but when $B_i = 0$, this inequality reduces to $\gamma> 3  (\mu_i \mu_i) / (\mu_j\alpha_j)^2$; the right-hand sides of these two inequalities differ by a factor of 4. Therefore, if $B_i$ and $\mu_i$ are oriented in the same direction in the Cartan subalgebra, there exists a window of values for $\gamma $ such that the system is in the condensed phase at $\mu = 0$, $B >0$ but returns to the uncondensed phase as $\mu$ increases. This is the same conclusion as what we found in the simpler $SU(2)$ case in the main text.

\end{document}